\documentclass[conference]{IEEEtran}

\IEEEoverridecommandlockouts                              
\normalsize
\usepackage[noadjust]{cite}
\usepackage{filecontents}
\usepackage{color,colortbl}
\usepackage[pdftex]{graphicx}
\usepackage{verbatim} 
\usepackage{epstopdf}
\usepackage[cmex10]{amsmath}
\usepackage[tight,footnotesize]{subfigure}
\usepackage{pifont}
\usepackage{algorithm}
\usepackage{algpascal}
\usepackage{algpseudocode}
\usepackage{multirow}
\usepackage{epstopdf}
\usepackage{microtype}

\definecolor{Gray}{gray}{0.99}
\newcommand{\E}{{\rm I\kern-.3em E}}

\title{Linearizing Active Antenna Arrays: \\ Digital Predistortion Method and Measurements}

\author{
\IEEEauthorblockN{Alberto Brihuega\IEEEauthorrefmark{1}, Mahmoud Abdelaziz\IEEEauthorrefmark{1},  Matias~Turunen\IEEEauthorrefmark{1}, Thomas Eriksson\IEEEauthorrefmark{2}, 
Lauri Anttila\IEEEauthorrefmark{1}, Mikko Valkama\IEEEauthorrefmark{1}
\IEEEauthorblockA{\IEEEauthorrefmark{1}Tampere University, Department of Electrical Engineering,
Tampere, Finland}
\IEEEauthorblockA{\IEEEauthorrefmark{2}Chalmers University of Technology, Department of Electrical Engineering,
Gothenburg, Sweden}
Email: alberto.brihuegagarcia@tuni.fi
}
\vspace{-7mm}

}

\begin{document}
       
\maketitle

\begin{abstract}
    In this paper, we provide a novel framework for efficient digital predistortion (DPD) based linearization of active antenna arrays with multiple and mutually different nonlinear power amplifiers. The proposed method builds on the use of a combined feedback signal essentially characterizing the observable nonlinear distortion at the receiving end. The proposed method is validated with extensive over-the-air RF measurements on a 64-element active antenna array transmitter operating at 28 GHz carrier frequency and transmitting a 200 MHz wide 5G New Radio (NR) waveform. The obtained results demonstrate the excellent linearization capabilities of the proposed solution, which allows for a very efficient implementation in practical systems.
    
\end{abstract}


\section{Introduction}

Power-efficient operation of transmitters is of fundamental importance in any modern wireless system, and is also one of the key design criteria for 5G NR base stations (BS) due to the large amount of antenna units that will be deployed \cite{intro_1}. Consequently, highly nonlinear power amplifiers (PAs) operating close to saturation are expected to be utilized, which, in turn, gives rise to high levels of nonlinear distortion \cite{GreenComm,PA_design}. 
Such nonlinearities induce inband and out-of-band distortions that may limit the capacity and quality-of-service of the intended and neighboring channel users. In order to compensate for the induced distortion, DPD solutions are generally adopted \cite{GMP_morgan,DPD_MM_5,DPD_DigitalMIMO}, consisting of the inverse nonlinearity of that exhibited by the PA, such that the resulting cascaded system is linearized. 

DPD-based linearization of single or few antenna transmitters is well established, however, the linearization of large antenna array transmitters is an open and challenging research problem \cite{Swedes_review}. In digital MIMO transmitters, deploying a dedicated DPD per transmitter or antenna unit may be unfeasible from the complexity point of view \cite{DPD_DigitalMIMO}. On the other hand, in analog or hybrid beamforming architectures, a single DPD unit must be able to linearize all the power amplifiers within the (sub)array, which is an underdetermined problem, and may result in reduced linearization performance due to mutual differences of the PA units \cite{DPD_MM_5}. However, by exploiting the spatial characteristics of the distortion, it is possible to develop efficient linearization solutions. As it was exposed in \cite{OOB_Mollen}, the nonlinear distortion gets beamformed towards the very same directions as the linear signal, while in other directions it gets diluted due to less coherent propagation. It was further shown in \cite{DPD_MM_5,DPD_DigitalMIMO} that by considering the signal from the beamformed channel perspective for DPD parameter learning, the nonlinear distortion can be kept at a sufficient low level, in all spatial directions, as a result of the combined effect of DPD linearization and beamforming. 

\begin{figure}[t!]
\centering
\includegraphics[width=.95\linewidth]{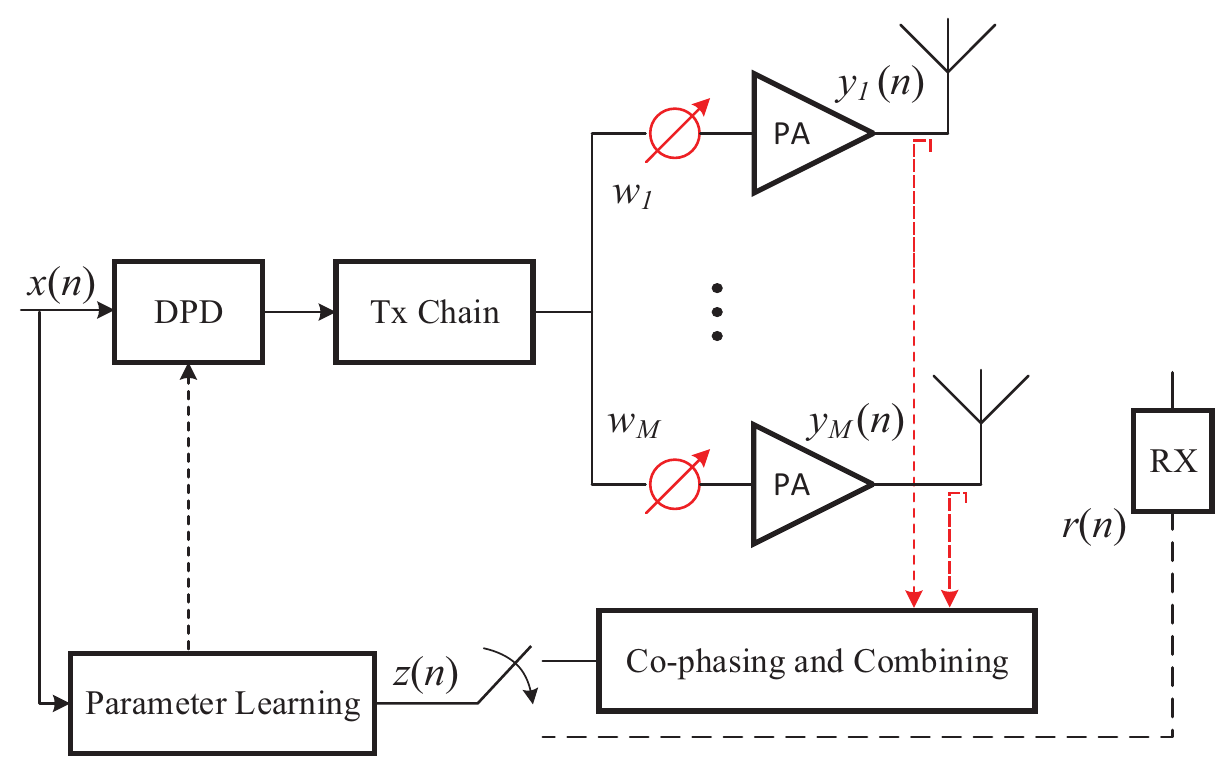}
\caption{Transmitter architecture and ILA-based DPD parameter learning based on combined feedback observation receiver or over-the-air feedback.}
\vspace{-3mm}
\label{fig:Tx}
\end{figure}

In this paper, we describe a novel DPD architecture in the context of active antenna arrays, illustrated in Fig. \ref{fig:Tx}, building on the combined feedback signal based learning. We provide over-the-air (OTA) measurements of the developed DPD solution with a 64-element active antenna array transmitter operating at 28 GHz carrier frequency and transmitting a 5G NR carrier with 200 MHz instantaneous bandwidth, representing a practical scenario of 5G NR deployment at the frequency range 2 (FR2) \cite{3GPPTS38104}. The obtained OTA measurements demonstrate very efficient linearization of the active array, despite adopting only a single DPD unit for the whole array. To our knowledge, these results are among the first demonstrating successful linearization of state-of-the-art active antenna array at mmWaves through actual measurements. 



\section{DPD Linearization Principle} \label{DPD_principle}
To shortly address the nonlinear distortion modeling in active array context, assume first that the DPD unit is off, and let $x_l(n) = w_lx(n)$ denote the I/Q samples of the $l$th PA input signal, with $w_l$ denoting the $l$th beamforming coefficient and $x(n)$ the complex baseband transmit signal. The corresponding PA output signal, assuming a generalized memory polynomial (GMP) PA model \cite{GMP_morgan} and $|w_l| = 1$, is given by 
\begin{equation}
\begin{split}
    y_l(n) &= w_l  \sum_{\substack{p=1 \\ p \text{odd}}}^{P} \sum_{\substack{g=-G}}^{G} \alpha_{l,p,g}(n) \star \left[|{x}(n-g)|^{p-1} x(n) \right].
\end{split}
\end{equation}
Here, $\star$ denotes the convolution operator, while $\alpha_{l,p,g}(n)$ is the impulse response of the $l$th PA for given nonlinear distortion order $p$ and envelope delay $g$. Furthermore, $P$ and $G$ denote the maximum considered nonlinearity order and leading/lagging envelope memory depth, respectively, while all impulse responses $\alpha_{l,p,g}(n)$ are assumed to be of length $M$. 
Assuming then, for presentation simplicity, that the beamforming coefficients are matched to the dominant phases of the array propagation channel, the combined signal at the receiving end is given by  
\begin{equation}
\begin{split}
    r(n) &= \sum_{\substack{l=1}}^{L}w_l^*y_l(n) \\
    &= \sum_{\substack{p=1 \\ p \text{odd}}}^{P} \sum_{\substack{g=-G}}^{G} \Bar{\alpha}_{p,g}(n) \star \left[|{x}(n-g)|^{p-1} {x}(n) \right],
\end{split}
\end{equation}
where $\Bar{\alpha}_{p,g}(n)  = \sum_{\substack{l=1}}^{L}\alpha_{l,p,g}(n)$ are the equivalent impulse responses of the array. This shows that the 
observable nonlinear distortion at the receiving end 
can be modeled with a single GMP.

Based on above modeling, we adopt a single GMP-based DPD unit with the purpose of linearizing the array from the received combined signal perspective. The GMP parameter learning builds then on mimicking or generating a local replica of the true received signal in (2), denoted as $z(n)$, which captures the nonlinear distortion stemming from the active array transmitter from the beamformed channel perspective. As shown in Fig. \ref{fig:Tx}, such feedback or observation signal for the parameter learning can be obtained by phase-aligning and combining the PA output signals. Alternatively, one could also argue for sending OTA feedback information or measurements from a true remote test receiver \cite{Swedes_review}. The GMP-based DPD coefficients, denoted by ${\beta}_{p,g}(n)$, are then learned by adopting an indirect learning architecture (ILA), where the feedback signal $z(n)$ is fed into a postdistorter whose coefficients are estimated by means of block least-squares, written as
\begin{equation}
    \boldsymbol{\beta} = (\mathbf{Z}^H\mathbf{Z})^{-1}\mathbf{Z}\mathbf{x}
\end{equation}
where $\mathbf{Z}$ is a matrix containing the set of GMP basis functions of the form $|{z}(n-g)|^{p-1} z(n)$, while $\mathbf{x}$ contains $N$ samples of the ideal baseband signal $x(n)$. Then, the postdistorter is applied as the actual predistorter. Despite ILA is adopted in this work, the proposed solution is also compatible with closed-loop learning  \cite{DPD_MM_5,DPD_DigitalMIMO} and direct learning, which are issues that we will focus on in our future extended work.


\section{Over-the-Air Measurement Results}\label{Measurements}
In this section, we provide extensive RF measurements on a 64-element active antenna array operating at 28 GHz carrier frequency in order to demonstrate the operation of the proposed DPD solution in timely 5G NR FR2 context.
\subsection{mmWave Measurement Setup}
\begin{figure}[t!]
\centering
\includegraphics[width=1\linewidth]{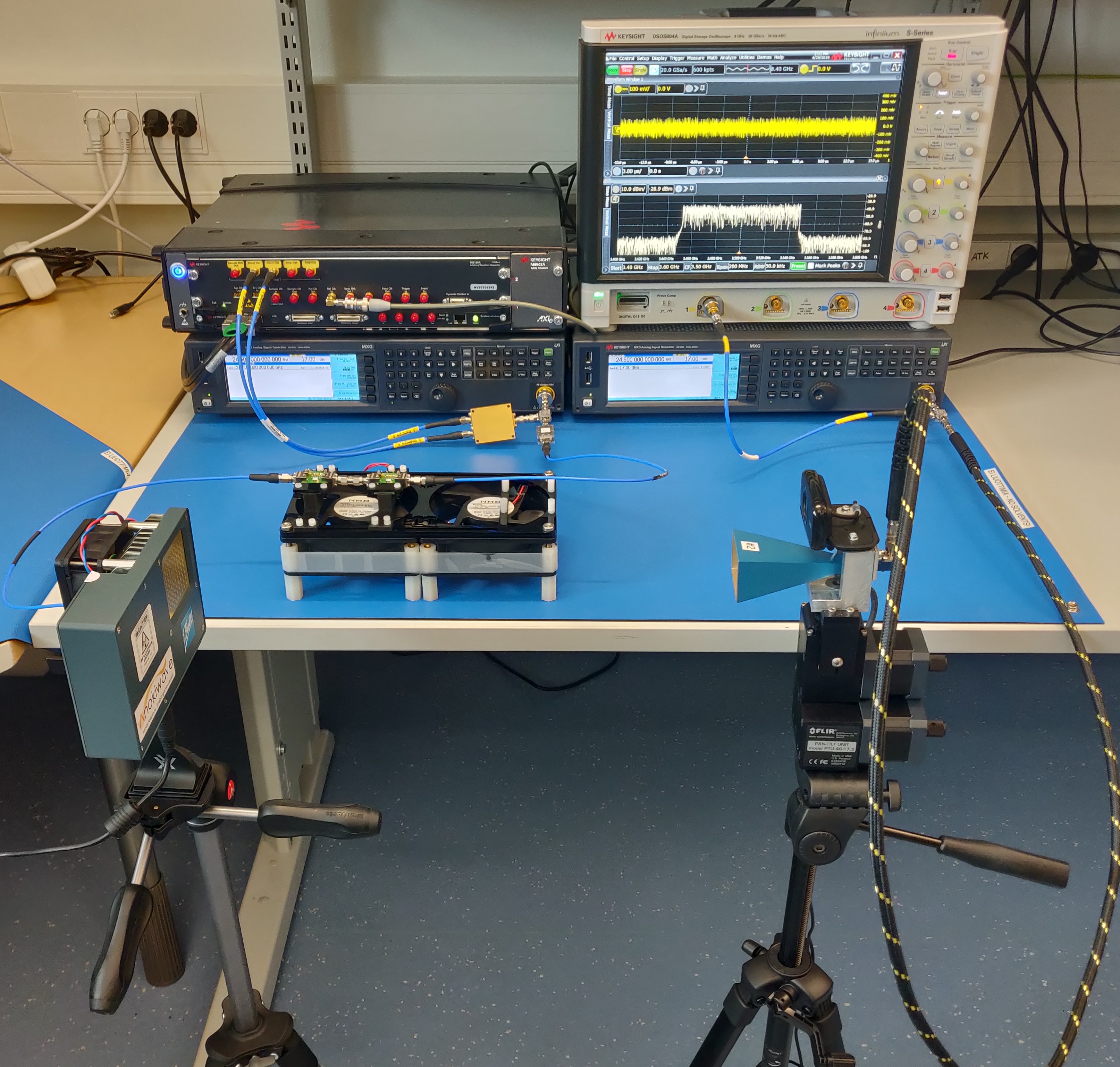}
\caption{mmWave OTA laboratory measurement setup.}
\label{fig:setup}
\end{figure}

The measurement setup is shown in Fig. \ref{fig:setup}. The Keysight M8190A arbitrary waveform generator is utilized to generate the modulated IF signal centered at 3.5 GHz. Two Keysight N5183B-MXG signal generators running at 24.5 GHz are used to generate the local oscillator signals that are employed to up-convert the IF signal to the desired carrier frequency, and then to downconvert it back to IF at the receiver side. The modulated RF waveform at 28 GHz is amplified with two HMC499LC4 driver amplifiers with 17 dB gain each that allow to feed the Anokiwave AWMF-0129 active antenna array with an adequate power level, such  that its PAs are driven close to saturation. The transmit signal then propagates over-the-air and is captured by a horn-antenna. After IF downconversion, the signal is captured by the Keysight DSOS804A oscilloscope that is utilized as the receiver/digitizer, and taken to baseband. The received samples are then processed in a host PC running \textsc{Matlab}, where the actual DPD learning and predistortion are performed, with oversampling factor of 4 wrt. the critical sample rate. 
%
In this setup, we utilize the actual OTA received signal to learn the DPD filter coefficients.

\subsection{Measurement Results and Analysis}
\begin{figure}[t!]
\centering
\includegraphics[width=.9\linewidth]{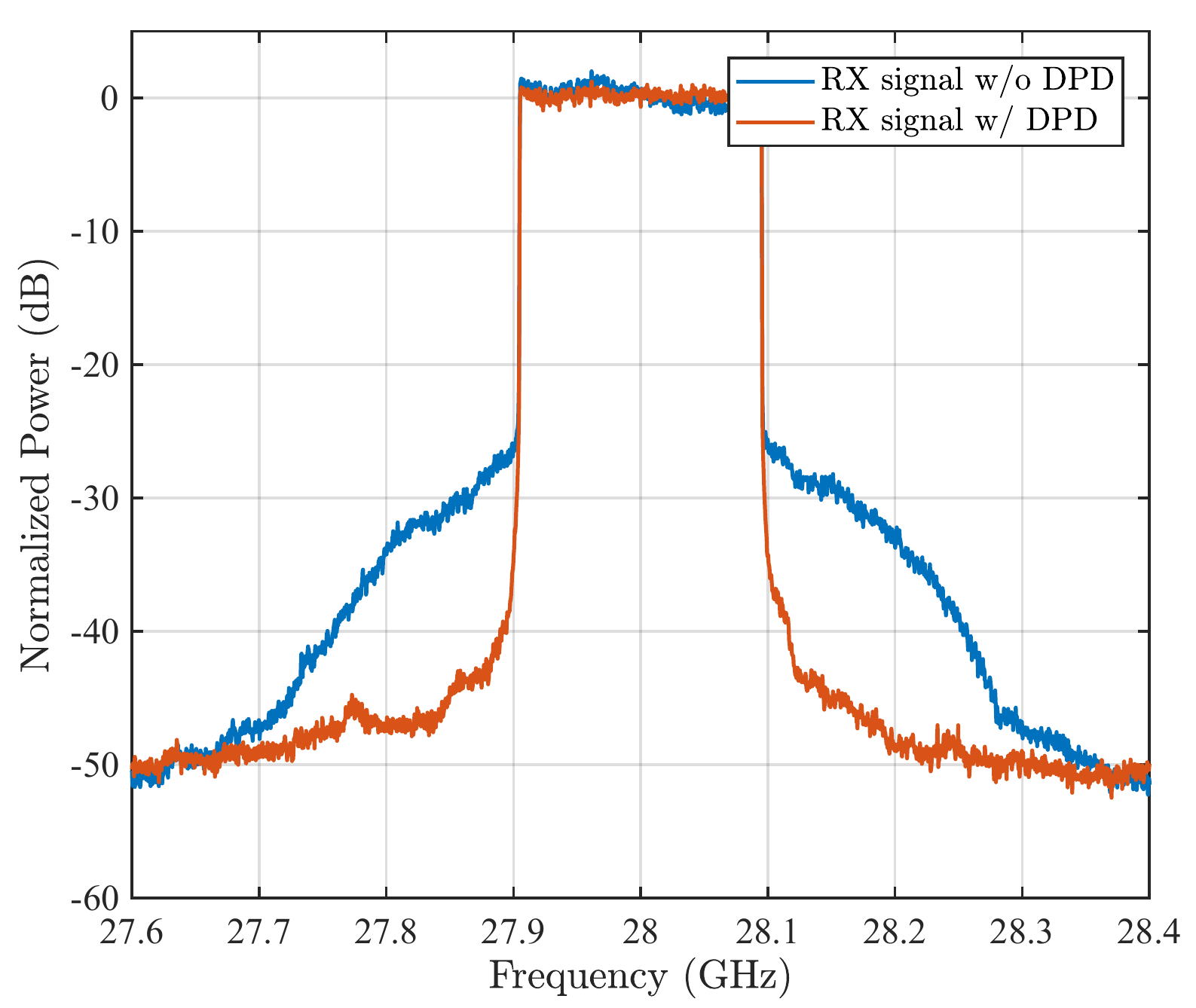}
\vspace{-2.8mm}
\caption{Combined received signal spectra at 28 GHz with and without the proposed DPD solution, adopting 5G NR waveform with 200 MHz carrier bandwidth and Anokiwave AWMF-0129 active antenna array. The passband powers are normalized to 0 dB, while true EIRP is ca. +38.5~dBm.}
\label{fig:RX_spectra}
\end{figure}

The power spectra of the OTA combined received signal, with and without  the proposed DPD, are shown in Fig. \ref{fig:RX_spectra}. The adopted test signal is a 200 MHz OFDM waveform with 64-QAM subcarrier modulation conforming to 3GPP 5G NR specifications \cite{3GPPTS38104}, with OFDM subcarrier spacing of 60 kHz, $K_{\mathrm{ACT}} = 3168$ active subcarriers and FFT size of $K_{\mathrm{FFT}}=4096$. For DPD learning, we consider three iterations of ILA utilizing LS-based parameter learning in each ILA run, as shown in (3). 
Each of the ILA iterations employs 40$k$ samples. The GMP-based DPD is parameterized as $P =7$, $M = 4$ and $G = 3$. The beam of the active antenna array is pointed towards 0 degrees, where the test receiver is located. 

\begin{figure}[t!]
\centering
\includegraphics[width=1\linewidth]{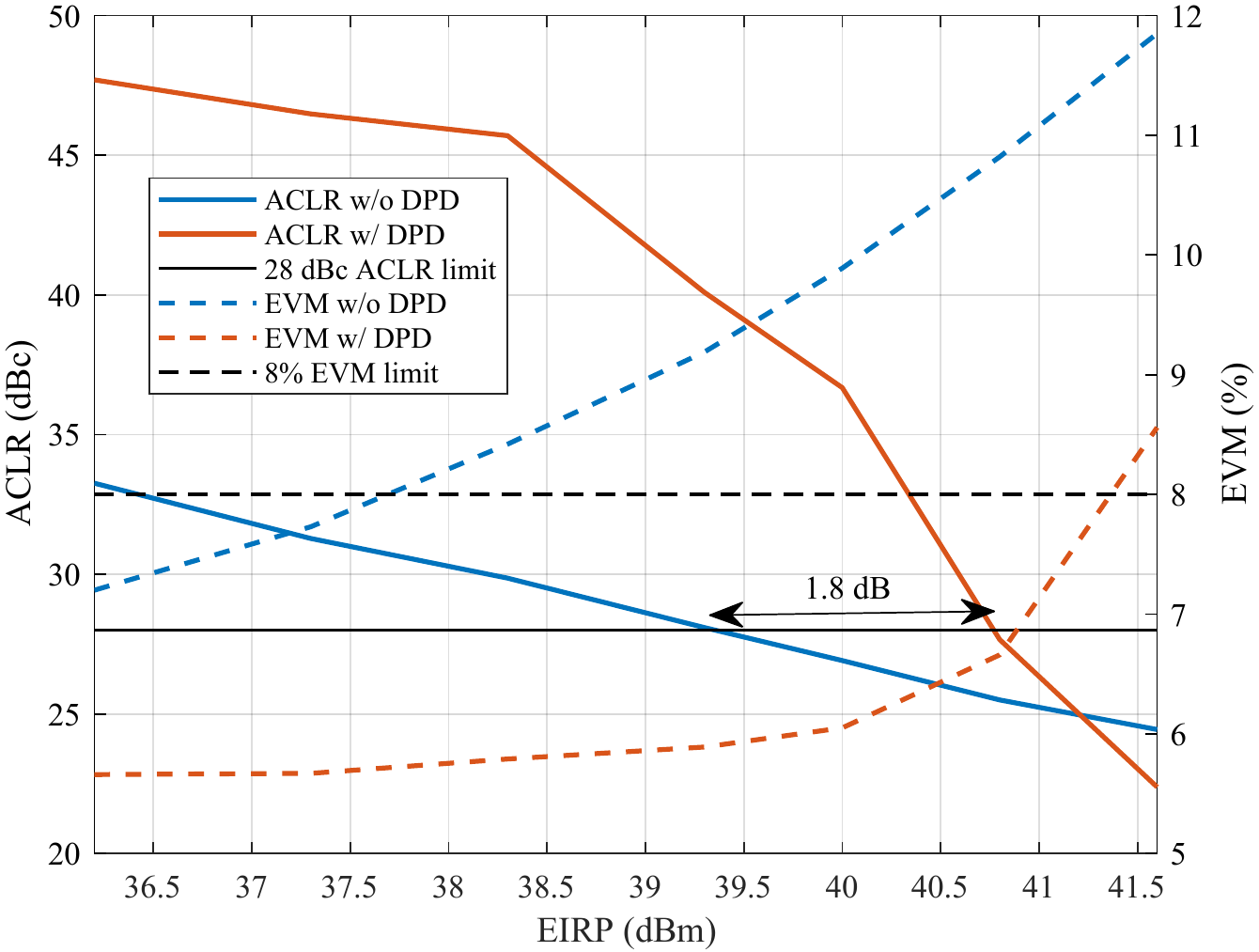}
\vspace{-7mm}
\caption{ACLR and EVM performance as a function of the EIRP at 28~GHz, when adopting 5G NR waveform with 200 MHz carrier bandwidth.}
\label{fig:power_sweep}
\end{figure}

As it can be observed in Fig. \ref{fig:RX_spectra}, excellent linearization is achieved 
when the proposed DPD is adopted. The resulting EVM and ACLR values are gathered in Table \ref{tab:EVM_ACLR}, which also shows the performance of a memory polynomial DPD with parameters $P =7$, $M = 4$ and $G = 0$, providing also great linearization of the transmitter despite its reduced complexity. In Table \ref{tab:EVM_ACLR}, the so-called \textit{active array input} refers to measuring the EVM and ACLR characteristics of the active array input directly, serving thus as a reference. Lastly, Fig. \ref{fig:power_sweep} shows the ACLR and EVM performance of the transmitter as a function of the effective isotropic radiated power (EIRP). The DPD performance decays very rapidly when the active antenna array is operated deeper into saturation, and even violating the 28 dBc ACLR and 8\% EVM limits specified for FR2 \cite{3GPPTS38104}. The adoption of the proposed DPD allows for close to 2 dB extra power  while fulfilling the emissions mask. Overall, these results demonstrate that an active array with very nonlinear PA units can be successfully linearized, building on the combined observed signal in parameter learning. 

\section{Conclusions and Future Work}\label{Conclusions}
In this paper, a digital predistortion framework for efficient linearization of active antenna array transmitters was presented. The proposed DPD solution was experimented and validated with practical over-the-air RF measurements utilizing a 64-element active array transmitter operating at 28 GHz and incorporating 5G NR waveform with 200 MHz carrier bandwidth. The obtained results show that a single DPD unit building on combined observed or feedback signal can efficiently linearize a bank of nonlinear PA units. 
These results are among the first in the literature, where active antenna array linearization is demonstrated at mmWaves through practical 5G NR compliant measurements and waveforms. 
Our future work will focus on more extensive quantitative measurements, particularly dealing with the analysis of the spatial behavior of the OOB emissions stemming from the active array transmitter. Furthermore, extending the DPD main path processing and parameter learning solutions to facilitate deeper compression of the PA units, as well as extending the parameter learning to closed-loop solutions, are of special interest.

\begin{table}[t!]
\caption{Measured EVM and ACLR Results at +38.5~dBm EIRP}
\centering
{\begin{tabular}{lll}\hline
    &\vline\:\:\: EVM ($\%$)  &\vline\:\:\: ACLR (dBc)    \\\hline
Active array input (for reference)                     &\vline\:\:\:  4.13      &\vline\:\:\:  50.32   \\\hline
Without DPD 				  &\vline\:\:\:    8.51     &\vline\:\:\:  29.86      \\\hline
With proposed DPD (MP / GMP) 							  &\vline\:\:\:       5.6\: / \:4.73  &\vline\:\:\: 43.5\: / \:45.7     \\\hline
\end{tabular}}{}
\label{tab:EVM_ACLR}
\end{table}


\bibliographystyle{IEEEbib}
\bibliography{Ref}
\end{document}